\documentclass[preprint,sort&compress,11pt,3p]{elsarticle}
\usepackage{moreverb}
\usepackage{mathtools}
\usepackage{amsmath}
\usepackage{amssymb}
\usepackage{algorithmic}
\usepackage{graphics}
\usepackage{graphicx}
\usepackage{subfigure}
\usepackage{caption}
\usepackage{enumitem}
\usepackage{extarrows}
\usepackage{color}
\usepackage{framed}
\usepackage{wrapfig}
\usepackage{bm}
\usepackage{mathrsfs}
\usepackage{mathabx}
\usepackage{multirow}
\usepackage{longtable}
\usepackage{hyperref}
\usepackage{paralist}
\usepackage{indentfirst}
\usepackage{relsize}
\usepackage{extarrows}
\usepackage{lineno,xcolor}
\usepackage{upgreek}
\usepackage{bm}
\usepackage{mwe}
\usepackage{xcolor}
\usepackage{booktabs}
\usepackage[flushleft]{threeparttable}
\usepackage[linesnumbered,lined,ruled]{algorithm2e}

\graphicspath{ {./Figure/} }

\newcommand{\eref}[1]{(\ref{#1})}

\hypersetup{
bookmarks=true,
bookmarksopen=true,
bookmarksnumbered=true,
unicode=false,
pdftoolbar=true,
pdfmenubar=true,
pdffitwindow=false,
pdfstartview={FitH},
pdftitle={My title},
pdfauthor={Author},
pdfsubject={Subject},
pdfcreator={Creator},
pdfproducer={Producer},
pdfkeywords={keywords},
pdfnewwindow=true,
colorlinks=true,
linkcolor=blue,
citecolor=blue,
filecolor=magenta,
urlcolor=blue
}

\journal{Journal name}

\begin{document}
	
\title{\Large Physics-informed neural network for seismic wave inversion in layered semi-infinite domain}

\author[NU1]{Pu Ren\fnref{equal}}
\author[NU2]{Chengping Rao\fnref{equal}}
\author[RUC]{Hao Sun}
\author[UCAS]{Yang Liu\corref{cor}}
\ead{liuyang22@ucas.ac.cn}
\cortext[cor]{Corresponding author.}
\fntext[equal]{Equal contribution.}

\address[NU1]{Department of Civil and Environmental Engineering, Northeastern University, Boston, MA 02115, USA}
\address[NU2]{Department of Mechanical and Industrial Engineering, Northeastern University, Boston, MA 02115, USA}
\address[RUC]{Gaoling School of Artificial Intelligence, Renmin University of China, Beijing, 100872, China}
\address[UCAS]{School of Engineering Sciences, University of Chinese Academy of Sciences, Beijing, 101408, China}

\begin{abstract}
\small
Estimating the material distribution of Earth's subsurface is a challenging task in seismology and earthquake engineering. The recent development of physics-informed neural network (PINN) has shed new light on seismic inversion. In this paper, we present a PINN framework for seismic wave inversion in layered (1D) semi-infinite domain. The absorbing boundary condition is incorporated into the network as a soft regularizer for avoiding excessive computation. In specific, we design a lightweight network to learn the unknown material distribution and a deep neural network to approximate solution variables. The entire network is end-to-end and constrained by both sparse measurement data and the underlying physical laws (i.e., governing equations and initial/boundary conditions). Various experiments have been conducted to validate the effectiveness of our proposed approach for inverse modeling of seismic wave propagation in 1D semi-infinite domain. 
\end{abstract}


\maketitle

\section{Introduction}\label{S:1}
Seismic imaging is essential to a wide range of scientific and engineering tasks. Thanks to the development of a theoretical foundation for wave propagation, researchers have been dedicated to tackling seismic inversion problems by combining physical simulation and field measurement data. One of the most important applications is full waveform inversion (FWI), which determines the high-resolution and high-fidelity velocity model (or material distribution) of the subsurface from the seismic waveforms collected on the field surface. This task is usually solved as an inverse problem of minimizing the misfit error between the observed seismic signals and the synthetic solution simulated based on an estimated velocity model. However, the traditional numerical methods suffer from three significant limitations for inverse analysis (e.g., FWI). First, there is a harsh requirement of computational resources and memory for calculating the Fr\'chet derivatives in the optimization procedure. Second, it limits the resolution of the inverted velocity model by using specific smoothing regularization approaches for alleviating the ill-posedness of inverse problems. The last one is that the final solution heavily depends on the initial guess of the model.

Due to the great advancement in machine learning (ML) and artificial intelligence (AI), scientists have embraced a new research direction for forward and inverse analysis of complex systems in the domain of seismology. For instance, deep neural networks (DNNs) and auto differentiation are utilized for solving the Eikonal equation~\cite{smith2020eikonet} and seismic inversion~\cite{zhu2021general,zhu2022integrating}. \cite{yang2021seismic} also explore the potential of Fourier Neural Operator~\cite{li2020fourier} for acoustic wave propagation and inversion. In addition to inferring velocity structures, \cite{gao2021deepgem} develop a generalized Expectation-Maximization (DeepGEM) algorithm for estimating the wave sources simultaneously. Nevertheless, these methods rely on large amounts of high-fidelity data, which are usually inaccessible in real-world applications. Therefore, there is an urgent need to develop novel ML approaches to address the sparsity and noise issues of measurement data. It is worthwhile to mention that Graph Network (GN)~\cite{battaglia2018relational} has been proven to be an effective tool for estimating the underlying velocity model and the corresponding initial condition of the specific wave propagation based on sparse measurement data~\cite{zhao2022learning}.

Moreover, the recent physics-informed neural networks (PINNs)~\cite{raissi2019physics,karniadakis2021physics} have also succeeded in forward and inverse analysis of various complex systems with limited sensor measurements~\cite{rao2020physics,raissi2020hidden,lu2021deepxde,haghighat2021physics,ren2022physics} or even no labeled data~\cite{sun2020surrogate,rao2021physics,gao2021phygeonet,ren2022phycrnet,gao2022physics}. The general principle of PINNs is to combine the known physical laws and the measured data to optimize the DNNs simultaneously. In particular, we observe a growing interest in applying PINNs for seismic wave modeling~\cite{bin2021pinneik,song2021solving,huang2022pinnup,rasht2022physics}, which can overcome the obstacles of traditional methods. Note that the current PINN implementations for seismic wave propagation primarily focus on studying acoustic wave propagation. 

In this paper, we aim to design PINN frameworks for elastic wave modeling and inversion in 1D domain, which is inspired by the previous work \cite{ren2022seismicnet}. To be more concrete, we first design a specific PINN architecture to approximate the solution variables of interest. Next, for the sake of seismic inversion, we introduce a new part of shallow NN to learn the material distribution (i.e., Young's modulus). The synthesized networks work simultaneously to model the specific seismic wave propagation. Furthermore, the absorbing boundary condition is considered here to truncate the domain and avoid unnecessary computation, which works as a soft regularizer in the loss function to constrain the entire network. Finally, we evaluate the performance of our proposed PINN frameworks on various numerical experiments, including solving wave equations and FWI. More specifically, we consider different material configurations when validating the method.

The rest of the paper is organized as follows. The Methods Section formulates the seismic inversion task (i.e., FWI) and introduces the proposed PINN framework. The Experiments Section presents various numerical experiments for validating the performance of our proposed method on seismic wave inversion. Moreover, we conclude the entire paper in the Conclusions Section.

\section{Method}\label{s:method}
\subsection{Problem setup}\label{s:prob}
In this paper, we focus on the inverse analysis of 1D seismic wave propagation in elastic media. First of all, the momentum and constitutive equations of elasticity in 1D are defined as 
\begin{equation} 
\label{eq:elasticity} 
    \begin{aligned}
        \rho\frac{\partial u^2}{\partial t^2}&=\frac{\partial \sigma}{\partial x},\\
    E\frac{\partial u}{\partial x} &=\sigma,
    \end{aligned}
\end{equation}
where $t$ and $x$ are the time and spatial dimensions; $\sigma(t,x)$ and $u(t,x)$ are the stress and displacement fields, respectively. $E(x)$ denotes Young's modulus with respect to $x$. $\rho$ represents the density. The initial condition of the displacement field is set as $u(0,:)=0$. In addition, we apply the absorbing boundary condition for the truncated boundary, which is given by 
\begin{equation} 
\label{eq:abc} 
    \frac{\partial u}{\partial t}+\sqrt{\frac{E}{\rho}}\cdot\frac{\partial u}{\partial x}=0.
\end{equation}

Our specific goal is to design PINN frameworks to take the spatiotemporal coordinate $(x,t)$ as inputs and approximate the displacement and stress variables $\{u,\sigma\}$ based on Eq.~\eref{eq:elasticity} and Eq.~\eref{eq:abc}. For the FWI problem, we aim to reconstruct the velocity model (in the form of Young's modulus) of the subsurface given the displacement waveform collected on the surface (i.e., $x=0$).

\subsection{PINN framework}\label{s:pinn}
In this part, we introduce the strategy of combining DNNs and physical laws (i.e., governing equations and boundary conditions) for learning seismic wave propagation. Various network architectures have been explored in the area of scientific ML, such as fully-connected NNs~\cite{raissi2019physics}, convolutional NNs~\cite{ren2022phycrnet}, graph NNs~\cite{gao2022physicsGNN,li2020multipole}, and transformer~\cite{geneva2022transformers}. Considering the complexity of our task, we employ the fully-connected feed-forward NNs to represent the seismic dynamics in this paper. The common implementation of fully-connected NNs utilizes an input layer, multiple hidden layers, and an output layer. In specific, the connection between $(i-1)$-th layer and $i$-th layer is given by
\begin{equation}
    \label{eq:dnn}
    \mathbf{X}^i = \sigma \left(\mathbf{W}^i \mathbf{X}^{i-1} + \mathbf{b}^i \right),\;\; i \in [1,n+1],
\end{equation}
where $\mathbf{X}^{i-1}$ and $\mathbf{X}^{i}$ are input and output features at $i$-th layer; $\mathbf{W}^i$ and $\mathbf{b}^i$ denote the weight tensor and bias vector of $i$-th layer; $n$ refers to the number of hidden layers; $\sigma(\cdot)$ represents the nonlinear activation function (e.g., Hyperbolic Tangent Function).

The major difference between pure data-driven ML and physics-informed learning lies in the introduction of physical constraints apart from the traditional data loss. The physics constraints are achieved by constructing the residual loss between the predicted dynamics and the underlying true physics. In PINNs, automatic differentiation~\cite{baydin2018automatic} is applied to obtain the derivative terms of interest. More details about the specific design of PINN architecture for seismic inversion can be found in Section~\ref{s:seismic_inversion}.

\begin{figure}[h!]
	\centering
	\includegraphics[width=1.0\textwidth]{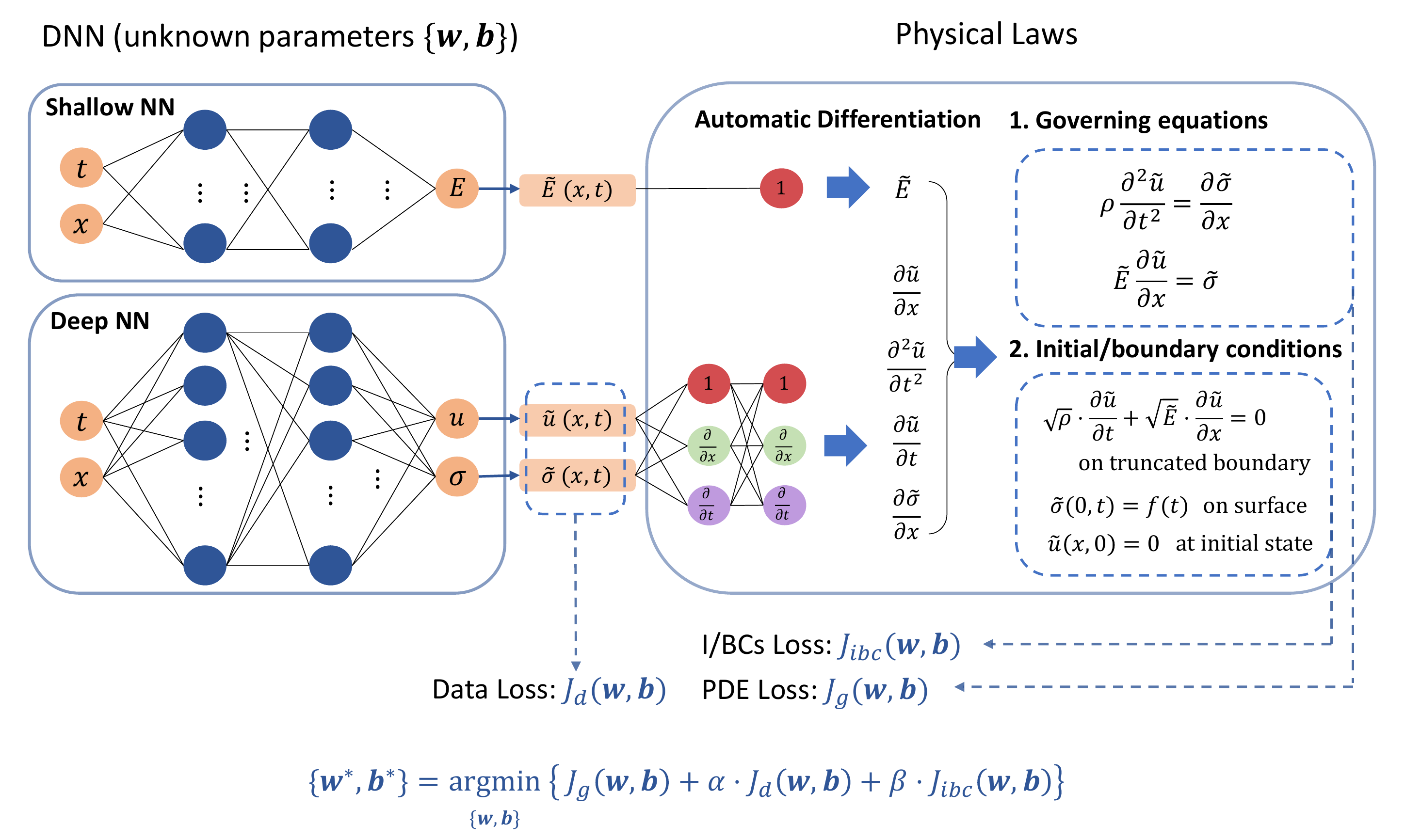}
	\caption{Architecture of the PINN for 1D full waveform inversion. A shallow neural network is used to approximate the unknown distribution of $E$.}
	\label{fig:FWI:PINN_architecture}
\end{figure}

\subsection{Seismic inversion}\label{s:seismic_inversion}
The goal of seismic inversion is to infer the high-fidelity velocity model (i.e., material distribution) of the subsurface based on the observed seismic signals (e.g., seismic waveforms), which are measured from the field surface. The challenge here is that the distribution of seismic wavespeed in the medium is unknown. To this end, we introduce another shallow NN to approximate the material variable $\widetilde{E}$ independently apart from the PINN architecture designed for approximating the solution variables $\widetilde{u}$ and $\widetilde{\sigma}$. The overall framework of our proposed method is presented in Fig.~\ref{fig:FWI:PINN_architecture}. Note that the shallow NN part, which simulates the material distribution, holds a smaller network size than the deep NN module due to less spatiotemporal complexity~\cite{haghighat2021physics,rasht2022physics} compared with learning solution variables. Herein, this new synergetic PINN framework serves as a model pipeline for FWI.

In addition, to avoid unnecessary computation, we consider applying the absorbing boundary condition for handling the semi-infinite domain in seismic wave propagation. The truncated boundary condition is defined as Eq.~\eref{eq:abc}. Finally, the loss function for inferring the velocity model is defined as
\begin{equation} 
\label{lossfunpde1} 
    J = J_g + \alpha\cdot J_d + \beta\cdot J_{ibc}
\end{equation}
where $\alpha$ and $\beta$ denote weighting coefficients. $J_g$, $J_{ibc}$, and $J_d$ denote the governing equation loss, initial/boundary condition loss, and data loss, respectively. The loss function guarantees the optimization of both the misfit between the predicted and the measured waveform, as well as the physical consistency described by Eq.~\eref{eq:elasticity}. Moreover, as shown in Fig.~\ref{fig:FWI:PINN_architecture}, Young's modulus $\widetilde{E}$ is approximated by the shallow NNs part, which is constrained by $J_g$ and $J_{ibc}$.

\section{Experiments}\label{s:experiment}
To validate the effectiveness of our proposed method, we conduct two numerical experiments on evaluating the performance of FWI in 1D domain. Specifically, two velocity models are designed: (1) a mixed velocity model with both linear and Gaussian functions; (2) a four-layer material distribution. All of the experiments are coded with TensorFlow~\cite{abadi2016tensorflow} and implemented on an NVIDIA Tesla V100 GPU card (32G) in a standard workstation.

\subsection{Setup}\label{s:setup}
\begin{figure}[h!]
	\centering
	\includegraphics[width=0.55\textwidth]{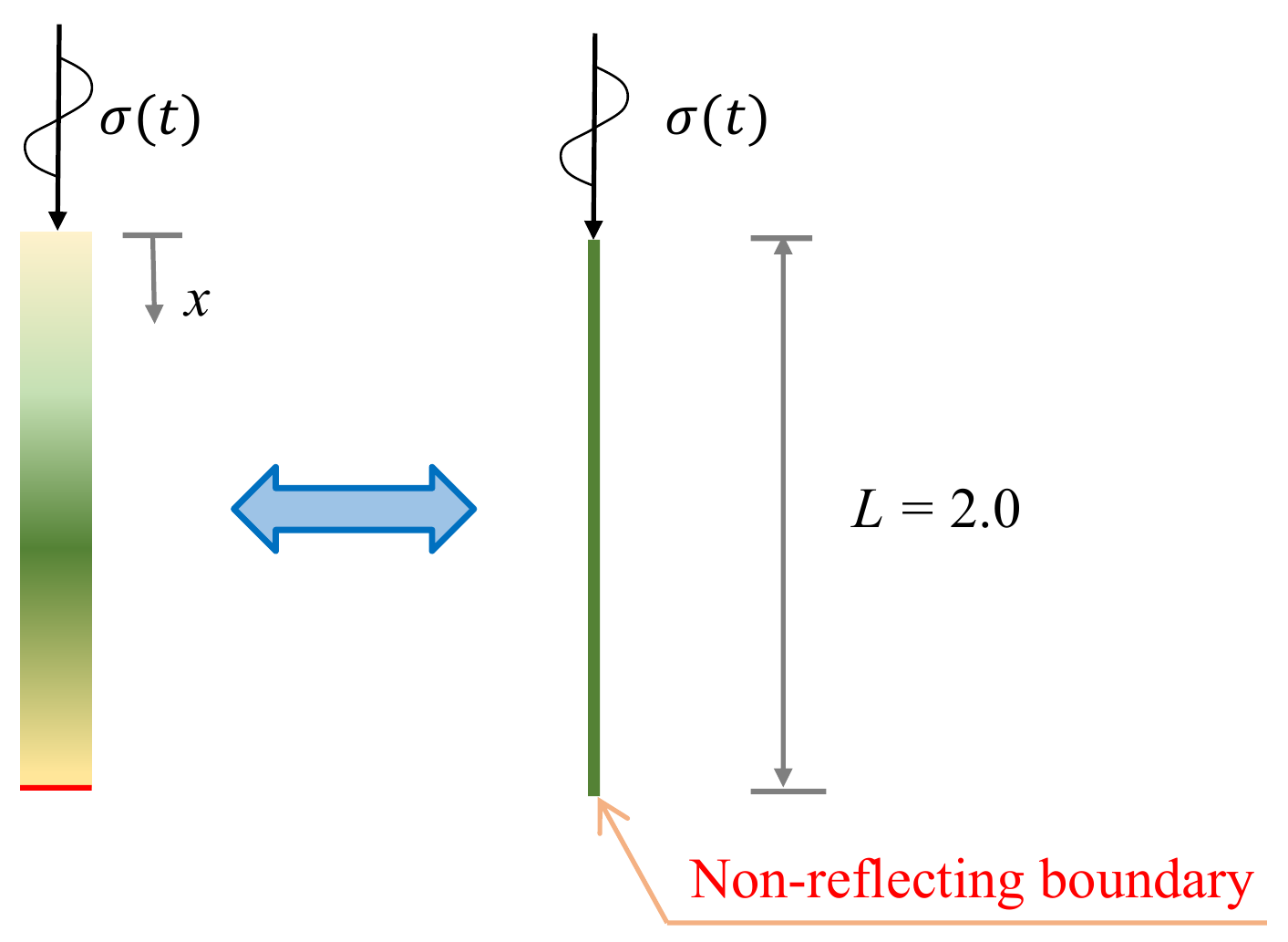}
	\caption{Computational domain of the 1D full waveform inversion problem. $F(t)$ denotes the time-varying load excited on the surface. The truncated depth is $L=2.0$.}
	\label{chap:FWI:diagram}
\end{figure}

In these numerical examples, we consider a 1D subsurface under the elasticity assumption, as shown in Fig.~\eref{chap:FWI:diagram}. To ensure the computational efficiency of PINN, we truncate the original computational semi-infinite domain with a finite depth, i.e., $x=2.0$. As a result, the absorbing boundary condition would be applied to the truncated boundary to avoid wave reflection. Note that the waveform we used for inversion is synthesized through finite element simulation in an enlarged domain $x\in[0, 10]$ though a truncated domain (i.e.,  $x\in[0, 2]$) is considered in PINN.

Furthermore, we also define an evaluation metric, the mean absolute relative error (MARE), to measure the prediction discrepancy of material distribution in the form of Young's modulus $E$. MARE is given by
\begin{equation}
    \label{eq:metric}
    \text{MARE} = \frac{1}{N} \sum_{i=1}^{N} \frac{|\widetilde{E}_i-E_i|}{|E_i|} \times 100\%,
\end{equation}
where $N$ is the number of collocation points.

\subsection{Linear and Gaussian material distribution}\label{s:gaussian_dist}
The first case considers a velocity model of the subsurface with both linear and Gaussian functions, which is defined in the form of Young's modulus 
\begin{equation}
    \label{eq:gaussian}
    E=2.0+0.9x+4.0\exp\left[-\frac{(x-1.1)^2}{0.2}\right].
\end{equation}
Moreover, in the simulation, we prescribe a time-varying force on the surface, which is given by 
\begin{equation} 
\label{fig:FWI:1D_pulse} 
    \sigma(t)=\sigma_0\cdot\exp\left[-\frac{(t-t_s)^2}{0.03}\right],
\end{equation}
where the amplitude $\sigma_0$ is set as 1.0 and the offset $t_s$ is 0.5. The entire simulation duration is 3.

We design the deep NNs of $50\times3$ (i.e., depth of 3 and width of 50) for approximating the solution variables $u$ and $\sigma$. The shallow NN has a network size of $4\times2$ (i.e., depth of 2 and width of 4) for learning the material distribution $E$. In addition, the Hyperbolic Tangent activation function (i.e., Tanh) is utilized in the framework. To train the composite PINN jointly, we adopt a routine with 2,000 iterations of Adam optimizer~\cite{kingma2014adam} followed by 40,000 iterations of L-BFGS-B optimizer~\cite{zhu1997algorithm}.

\begin{figure}[h!]
	\centering
	\includegraphics[width=0.6\textwidth]{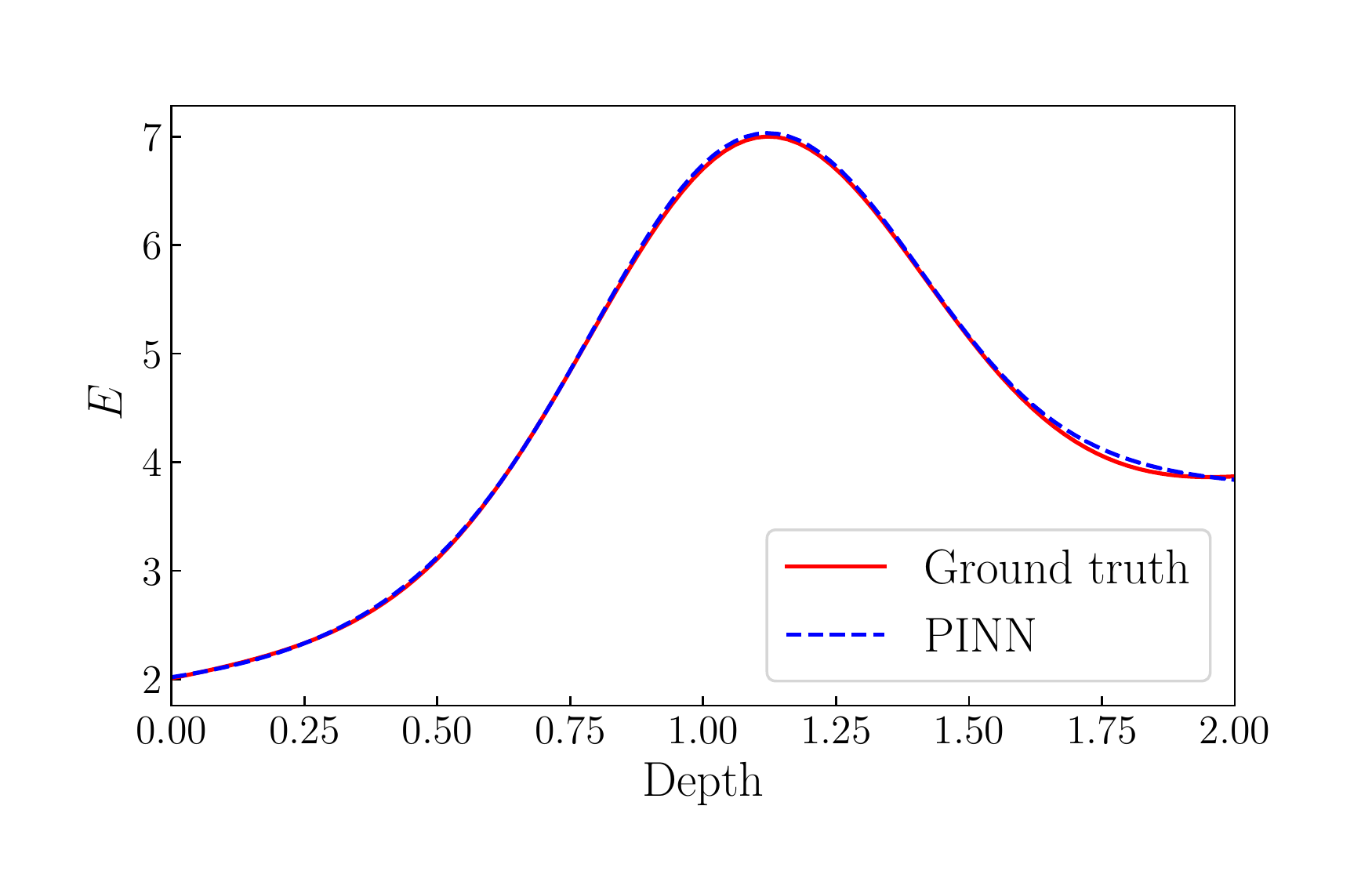}
	\caption{Comparison between the predicted material distribution in the form of Young's modulus $E$ and the true velocity model.}
	\label{fig:FWI:E_distribution}
\end{figure}

\begin{figure}[h!]
	\centering
	\includegraphics[width=0.6\textwidth]{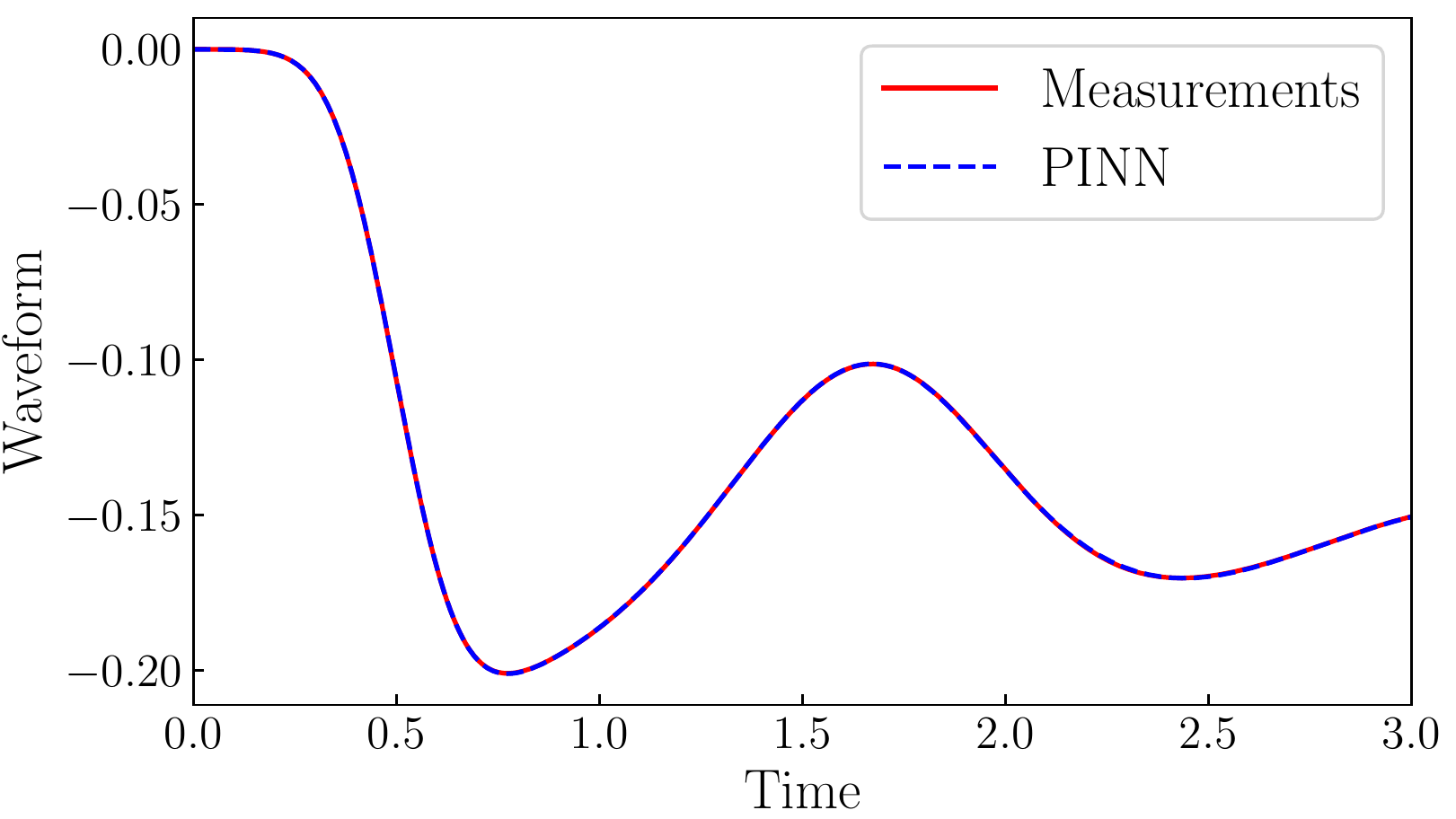}
	\caption{Comparison between the learned surface waveform (i.e., displacement) and the ground truth.}
	\label{fig:FWI:surf_pred_vs_meas}
\end{figure}

In addition, 30,000 collocation points are randomly sampled within the spatiotemporal domain for constructing the $J_g$. The synthetic wave signal includes 301 data points collected in $t\in[0,3]$. When the optimization convergences, we can directly infer the material distribution $E$ and predict the waveform on the surface. The comparison between the predicted velocity model and the ground truth is shown in Fig. \ref{fig:FWI:E_distribution}. It is obvious that the learned velocity model agrees well with the ground truth despite the minor discrepancy near the truncated boundary. In specific, the evaluation metric MARE shows that the prediction error of material distribution is only $0.47\%$. Furthermore, we present the comparison between the predicted surface waveform (i.e., displacement responses) and the synthetic measurements in Fig.~\ref{fig:FWI:surf_pred_vs_meas}. The result also proves an excellent performance of our proposed PINN framework in approximating the solution variables. 

\subsection{Four-layer material distribution}\label{s:4layer}
The second numerical example is a four-layer velocity model, which is a more complicated subsurface compared with the first case. The mathematical formulation is given by 
\begin{equation}
E = 
    \begin{cases} 
    1.5, & 0\le x\le 0.5 \\ 
    6.0, & 0.5\le x \le 1.0\\
    4.0, & 1.0 \le x\le 1.5 \\
    6.5, & \text{others} \\
    \end{cases}.
\end{equation}
To avoid the discontinuity of Young's modulus $E$, we smooth the transition between two adjacent layers with the Sine function. Note that the derivatives of $E$ are still discontinuous. The transition length is set to 0.1. This problem becomes more complicated since the distribution of $E$ would dissipate the seismic wave during propagation and lead to a faint waveform. Therefore, the excitation of surface load applied here features a higher frequency which is described by
\begin{equation} 
\label{fig:FWI:1D_pulse_HF} 
    \sigma(t)=\sigma_0\cdot\exp\left[-\frac{(t-t_s)^2}{0.008}\right],
\end{equation}
$\sigma_0$, $t_s$, and time duration are still defined as 1.0, 0.5, and 3, respectively.

\begin{figure}[t!]
\centering
\includegraphics[width=0.6\textwidth]{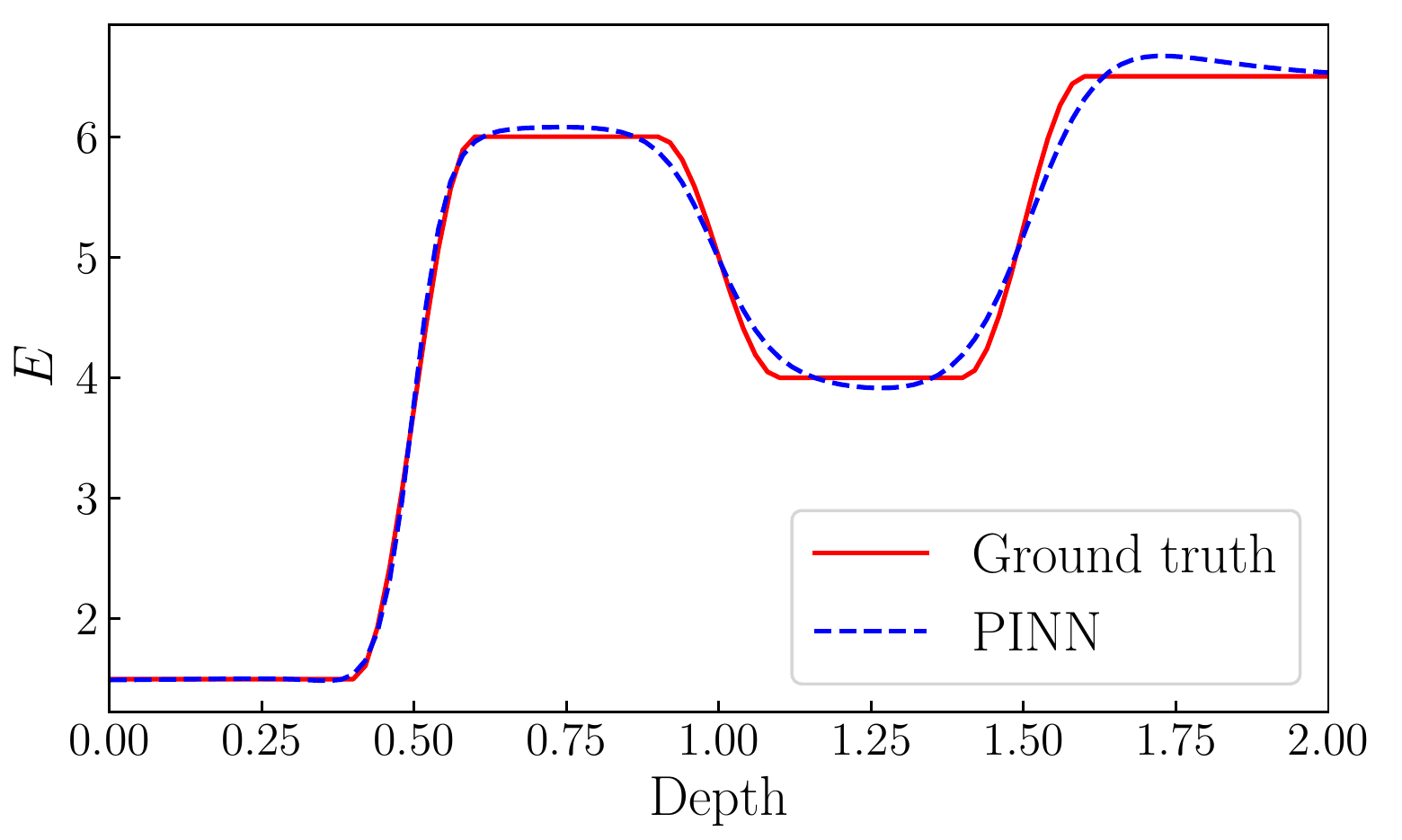}
\caption{Comparison between the predicted material distribution in the form of Young's modulus $E$ and the true velocity model.}
\label{fig:FWI:E_distribution_4_layer}
\end{figure}

The NNs with more powerful expressiveness are employed to approximate the solutions, i.e., deep NN with a network size of $80\times4$ and shallow NN with the size of $10\times2$ for learning $E$. Additionally, 100,000 collocation points are sampled within the spatiotemporal domain. 5,000 iterations of Adam optimizer and 400,000 iterations of L-BFGS-B optimizer are used to train the composite PINNs. The inverted distribution of $E$ is inferred from the trained model for comparison with the ground truth, as shown in Fig.~\ref{fig:FWI:E_distribution_4_layer}. It can be seen that the PINN is capable of identifying the subsurface with four layers. The prediction of four-layer material distribution has a MARE of $1.70\%$, which also exhibits great performance for the complex velocity model. Nevertheless, we also observe that the discrepancy increases as the depth increases. This is owing to the inherent nature of this problem -- as the wave propagates deep, the reflected wave that is observable on the surface carries fewer features. It reflects the poor identifiability of the deeper subsurface and the ill-posedness of such tasks. Moreover, the discontinuity (i.e., derivatives of $E$) near the interface of adjacent layers also makes it challenging to accurately reconstruct the subsurface material as the PINN is a continuous approximator. Besides, the comparison of the surface waveform between the prediction from PINN and the ground truth also validates the effectiveness of our proposed method. To summarize, our designed PINN framework shows promising results on seismic inversion problems. 

\begin{figure}[t!]
\centering
\includegraphics[width=0.6\textwidth]{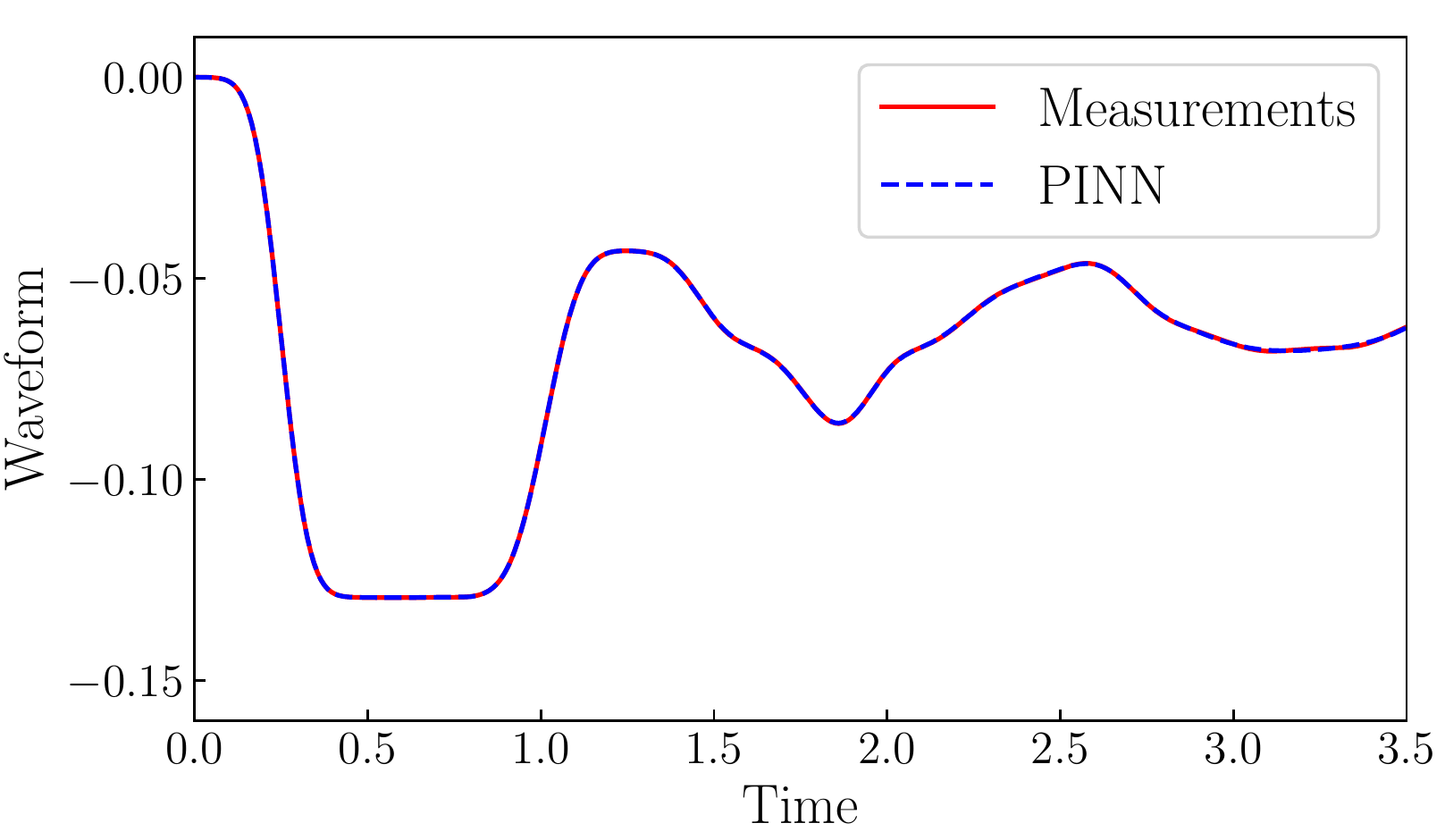}
\caption{Comparison between the learned surface waveform (i.e., displacement) and the ground truth.}
\label{fig:FWI:surf_pred_vs_meas_4_layer}
\end{figure}

\section{Conclusions}\label{s:conclusion}
This paper introduces a novel PINN framework for learning seismic wave propagation in elastic media. In specific, a deep NN is designed to approximate the physical laws (governing equations and the initial/boundary condition), and a shallow NN is employed to learn the underlying velocity structure. Moreover, the absorbing boundary is incorporated into the network to truncate the semi-infinite domain and avoid excessive computation. We validate the effectiveness of our proposed approach in two numerical cases in the context of seismic inversion (i.e., FWI). In the future, we would like to explore the potential of physics-informed learning for high-dimensional seismic inversion problems (e.g., in 2D/3D domains) but are not limited to employing fully-connected NNs that are typically used in PINNs. This is due to the observation that the physics-informed discrete learning scheme~\cite{ren2022phycrnet,ren2022physics,gao2021phygeonet} is promising to estimate more complex velocity models.


\section*{Acknowledgement}
P. Ren would like to gratefully acknowledge the Vilas Mujumdar Fellowship at Northeastern University. Y. Liu would like to thank the support from the Fundamental Research Funds for the Central Universities. The work is also supported by the National Natural Science Foundation of China (No. 62276269).

\bibliographystyle{elsarticle-num}
\bibliography{refs.bib}

\begin{thebibliography}{10}
\expandafter\ifx\csname url\endcsname\relax
  \def\url#1{\texttt{#1}}\fi
\expandafter\ifx\csname urlprefix\endcsname\relax\def\urlprefix{URL }\fi
\expandafter\ifx\csname href\endcsname\relax
  \def\href#1#2{#2} \def\path#1{#1}\fi

\bibitem{smith2020eikonet}
J.~D. Smith, K.~Azizzadenesheli, Z.~E. Ross, Eikonet: Solving the eikonal
  equation with deep neural networks, IEEE Transactions on Geoscience and
  Remote Sensing (2020).

\bibitem{zhu2021general}
W.~Zhu, K.~Xu, E.~Darve, G.~C. Beroza, A general approach to seismic inversion
  with automatic differentiation, Computers \& Geosciences 151 (2021) 104751.

\bibitem{zhu2022integrating}
W.~Zhu, K.~Xu, E.~Darve, B.~Biondi, G.~C. Beroza, Integrating deep neural
  networks with full-waveform inversion: Reparameterization, regularization,
  and uncertainty quantification, Geophysics 87~(1) (2022) R93--R109.

\bibitem{yang2021seismic}
Y.~Yang, A.~F. Gao, J.~C. Castellanos, Z.~E. Ross, K.~Azizzadenesheli, R.~W.
  Clayton, Seismic wave propagation and inversion with neural operators, The
  Seismic Record 1~(3) (2021) 126--134.

\bibitem{li2020fourier}
Z.~Li, N.~Kovachki, K.~Azizzadenesheli, B.~Liu, K.~Bhattacharya, A.~Stuart,
  A.~Anandkumar, Fourier neural operator for parametric partial differential
  equations, arXiv preprint arXiv:2010.08895 (2020).

\bibitem{gao2021deepgem}
A.~Gao, J.~Castellanos, Y.~Yue, Z.~Ross, K.~Bouman, Deepgem: Generalized
  expectation-maximization for blind inversion, Advances in Neural Information
  Processing Systems 34 (2021) 11592--11603.

\bibitem{battaglia2018relational}
P.~W. Battaglia, J.~B. Hamrick, V.~Bapst, A.~Sanchez-Gonzalez, V.~Zambaldi,
  M.~Malinowski, A.~Tacchetti, D.~Raposo, A.~Santoro, R.~Faulkner, et~al.,
  Relational inductive biases, deep learning, and graph networks, arXiv
  preprint arXiv:1806.01261 (2018).

\bibitem{zhao2022learning}
Q.~Zhao, D.~B. Lindell, G.~Wetzstein, Learning to solve pde-constrained inverse
  problems with graph networks, arXiv preprint arXiv:2206.00711 (2022).

\bibitem{raissi2019physics}
M.~Raissi, P.~Perdikaris, G.~E. Karniadakis, Physics-informed neural networks:
  A deep learning framework for solving forward and inverse problems involving
  nonlinear partial differential equations, Journal of Computational physics
  378 (2019) 686--707.

\bibitem{karniadakis2021physics}
G.~E. Karniadakis, I.~G. Kevrekidis, L.~Lu, P.~Perdikaris, S.~Wang, L.~Yang,
  Physics-informed machine learning, Nature Reviews Physics 3~(6) (2021)
  422--440.

\bibitem{rao2020physics}
C.~Rao, H.~Sun, Y.~Liu, Physics-informed deep learning for incompressible
  laminar flows, Theoretical and Applied Mechanics Letters 10~(3) (2020)
  207--212.

\bibitem{raissi2020hidden}
M.~Raissi, A.~Yazdani, G.~E. Karniadakis, Hidden fluid mechanics: Learning
  velocity and pressure fields from flow visualizations, Science 367~(6481)
  (2020) 1026--1030.

\bibitem{lu2021deepxde}
L.~Lu, X.~Meng, Z.~Mao, G.~E. Karniadakis, Deepxde: A deep learning library for
  solving differential equations, SIAM Review 63~(1) (2021) 208--228.

\bibitem{haghighat2021physics}
E.~Haghighat, M.~Raissi, A.~Moure, H.~Gomez, R.~Juanes, A physics-informed deep
  learning framework for inversion and surrogate modeling in solid mechanics,
  Computer Methods in Applied Mechanics and Engineering 379 (2021) 113741.

\bibitem{ren2022physics}
P.~Ren, C.~Rao, Y.~Liu, Z.~Ma, Q.~Wang, J.-X. Wang, H.~Sun, Physics-informed
  deep super-resolution for spatiotemporal data, arXiv preprint
  arXiv:2208.01462 (2022).

\bibitem{sun2020surrogate}
L.~Sun, H.~Gao, S.~Pan, J.-X. Wang, Surrogate modeling for fluid flows based on
  physics-constrained deep learning without simulation data, Computer Methods
  in Applied Mechanics and Engineering 361 (2020) 112732.

\bibitem{rao2021physics}
C.~Rao, H.~Sun, Y.~Liu, Physics-informed deep learning for computational
  elastodynamics without labeled data, Journal of Engineering Mechanics 147~(8)
  (2021) 04021043.

\bibitem{gao2021phygeonet}
H.~Gao, L.~Sun, J.-X. Wang, Phygeonet: Physics-informed geometry-adaptive
  convolutional neural networks for solving parameterized steady-state pdes on
  irregular domain, Journal of Computational Physics 428 (2021) 110079.

\bibitem{ren2022phycrnet}
P.~Ren, C.~Rao, Y.~Liu, J.-X. Wang, H.~Sun, Phycrnet: Physics-informed
  convolutional-recurrent network for solving spatiotemporal pdes, Computer
  Methods in Applied Mechanics and Engineering 389 (2022) 114399.

\bibitem{gao2022physics}
H.~Gao, M.~J. Zahr, J.-X. Wang, Physics-informed graph neural galerkin
  networks: A unified framework for solving pde-governed forward and inverse
  problems, Computer Methods in Applied Mechanics and Engineering 390 (2022)
  114502.

\bibitem{bin2021pinneik}
U.~bin Waheed, E.~Haghighat, T.~Alkhalifah, C.~Song, Q.~Hao, Pinneik: Eikonal
  solution using physics-informed neural networks, Computers \& Geosciences 155
  (2021) 104833.

\bibitem{song2021solving}
C.~Song, T.~Alkhalifah, U.~B. Waheed, Solving the frequency-domain acoustic vti
  wave equation using physics-informed neural networks, Geophysical Journal
  International 225~(2) (2021) 846--859.

\bibitem{huang2022pinnup}
X.~Huang, T.~Alkhalifah, Pinnup: Robust neural network wavefield solutions
  using frequency upscaling and neuron splitting, Journal of Geophysical
  Research: Solid Earth 127~(6) (2022) e2021JB023703.

\bibitem{rasht2022physics}
M.~Rasht-Behesht, C.~Huber, K.~Shukla, G.~E. Karniadakis, Physics-informed
  neural networks (pinns) for wave propagation and full waveform inversions,
  Journal of Geophysical Research: Solid Earth 127~(5) (2022) e2021JB023120.

\bibitem{ren2022seismicnet}
P.~Ren, C.~Rao, H.~Sun, Y.~Liu, Seismicnet: Physics-informed neural networks
  for seismic wave modeling in semi-infinite domain, arXiv preprint
  arXiv:2210.14044 (2022).

\bibitem{gao2022physicsGNN}
H.~Gao, M.~J. Zahr, J.-X. Wang, Physics-informed graph neural galerkin
  networks: A unified framework for solving pde-governed forward and inverse
  problems, Computer Methods in Applied Mechanics and Engineering 390 (2022)
  114502.

\bibitem{li2020multipole}
Z.~Li, N.~Kovachki, K.~Azizzadenesheli, B.~Liu, A.~Stuart, K.~Bhattacharya,
  A.~Anandkumar, Multipole graph neural operator for parametric partial
  differential equations, Advances in Neural Information Processing Systems 33
  (2020) 6755--6766.

\bibitem{geneva2022transformers}
N.~Geneva, N.~Zabaras, Transformers for modeling physical systems, Neural
  Networks 146 (2022) 272--289.

\bibitem{baydin2018automatic}
A.~G. Baydin, B.~A. Pearlmutter, A.~A. Radul, J.~M. Siskind, Automatic
  differentiation in machine learning: a survey, Journal of Marchine Learning
  Research 18 (2018) 1--43.

\bibitem{abadi2016tensorflow}
M.~Abadi, P.~Barham, J.~Chen, Z.~Chen, A.~Davis, J.~Dean, M.~Devin,
  S.~Ghemawat, G.~Irving, M.~Isard, et~al., ${TensorFlow}$: a system for
  ${Large-Scale}$ machine learning, in: 12th USENIX symposium on operating
  systems design and implementation (OSDI 16), 2016, pp. 265--283.

\bibitem{kingma2014adam}
D.~P. Kingma, J.~Ba, Adam: A method for stochastic optimization, arXiv preprint
  arXiv:1412.6980 (2014).

\bibitem{zhu1997algorithm}
C.~Zhu, R.~H. Byrd, P.~Lu, J.~Nocedal, Algorithm 778: L-bfgs-b: Fortran
  subroutines for large-scale bound-constrained optimization, ACM Transactions
  on mathematical software (TOMS) 23~(4) (1997) 550--560.

\end{thebibliography}

\end{document}